\documentclass[a4paper,12pt]{article}
\pdfoutput=1
\usepackage{amssymb}
\usepackage{amsmath}
\usepackage{epsfig}
\usepackage{graphicx}
\usepackage{color,ulem}
\usepackage{hyperref}

\makeatletter
\renewcommand{\theequation}{\thesection.\arabic{equation}}
\@addtoreset{equation}{section}
\makeatother
\def\be{\begin{equation}}
\def\ee{\end{equation}}
\def\bea{\begin{eqnarray}}
\def\eea{\end{eqnarray}}
\def\eq{\begin{eqnarray}}
\def\eqx{\end{eqnarray}}

\def\({\left(}
\def\){\right)}
\def\<{\left<}
\def\>{\right>}

\def\be{\begin{equation}}
\def\ee{\end{equation}}
\def\ben{\begin{eqnarray}}
\def\een{\end{eqnarray}}
\def\({\left(}
\def\){\right)}
\def\<{\left<}
\def\>{\right>}
\def\!{\right|}
\def\|{\left|}

\def\[{\left[}
\def\]{\right]}

\def\+{\bar}

\def\tu{{\tilde{u}}}
\def\tD{{\tilde{\Delta}}}
\def\nth{{\negthinspace}}

\usepackage{hyperref}
\hypersetup{colorlinks,%
  citecolor=blue,%
  linkcolor=blue,%
  pdftex}

\begin{document}

\begin{titlepage}
\vskip1cm
\begin{flushright}
\end{flushright}
\vskip0.25cm
\centerline{
\bf \Large 
Experimental Probes of  Traversable 
Wormholes
} 
\vskip1cm \centerline{ \textsc{
 Dongsu Bak,$^{\, \tt a,c}$  Chanju Kim,$^{\, \tt b}$ Sang-Heon Yi$^{\, \tt a}$} }
\vspace{1cm} 
\centerline{\sl  a) Physics Department,
University of Seoul, Seoul 02504 \rm KOREA}
 \vskip0.3cm
 \centerline{\sl b) Department of Physics, Ewha Womans University,
  Seoul 03760 \rm KOREA}
   \vskip0.3cm
 \centerline{\sl c) Natural Science Research Institute,
University of Seoul, 
Seoul 02504 \rm KOREA}
 \vskip0.4cm
 \centerline{
\tt{(dsbak@uos.ac.kr,\,cjkim@ewha.ac.kr,\,shyi704@uos.ac.kr})} 
  \vspace{2cm}
\centerline{ABSTRACT} \vspace{0.75cm} 
{
\noindent 
We propose possible probes which could be used to demonstrate experimentally 
the existence of the bulk and the formation of a traversable wormhole purely in
terms of boundary operations only. In the two-dimensional Einstein-dilaton
gravity, the traversable wormhole is realized by turning on a double trace
interaction which couples the two boundaries of the AdS$_2$ black hole.
Signals can propagate in the traversable wormhole through two different
channels. The boundary channel is direct and instantaneous, while
the bulk channel respects the bulk causality and takes a certain amount 
of time to complete signaling. In the latter case, we show that the signal
frequency detected on the other side is highly modulated in general. The
time delay as well as the frequency-modulation pattern could then be clear
indications that the signal comes out through the bulk channel. We discuss
the characteristics of the observed signal more explicitly for simple 
transitional configurations of the black hole from/to the eternal 
traversable wormhole.

}

\end{titlepage}


\section{Introduction}
Recently, there are remarkable developments on two-dimensional Einstein-dilaton gravity, dubbed as the Jackiw-Teitelboim(JT) model~\cite{Jackiw:1984je,Teitelboim:1983ux}. The absence of the bulk gravitational degree of freedom in this model lowers the computational and conceptual hurdles  and the resultant boundary theory, which is known as the Schwarzian theory~\cite{Jensen:2016pah,Engelsoy:2016xyb,Maldacena:2016upp,Maldacena:2017axo}, has led to full quantum computations, very recently~\cite{Stanford:2017thb,Saad:2019lba,Stanford:2019vob} (see~\cite{Muta:1992xw} in some other direction). In the context of the AdS/CFT correspondence,  the boundary dual theory is also identified with the Sachdev-Ye-Kitaev(SYK) model~\cite{SYK}, which spurs various works on expanding our understanding.  These developments give us various new insights in  quantum gravity, though those seem to be strongly model-dependent at this stage. However, since the JT model appears as a universal low energy limit of the near horizon on extremal black holes in the form of nearly AdS$_{2}$ or nearly dS$_{2}$ gravity, it would give us useful insights to the quantum nature of black holes and even in the quantum nature of cosmology~\cite{Maldacena:2019cbz,Cotler:2019nbi}.

In this note we shall be mainly concerned with bulk aspects of traversable wormholes  in  the JT model,  which requires the violation of the averaged null energy condition(ANEC) in a specific way~\cite{Gao:2016bin,Maldacena:2017axo,Maldacena:2018gjk}.  Although this specific  ANEC violating left/right(L/R) boundary  interaction or boundary sources are non-local and lead to a direct instantaneous interaction between L/R boundaries,  the bulk locality and causality seem to be preserved in some sense.  The usual bulk gravitational  interpretation of traversable wormholes in our model is completely consistent and provides a powerful tool for the probe to the GR=QM picture~\cite{Susskind:2018pmk,Susskind:2017ney,Maldacena:2013xja}. Since the bulk causality seems crucial  in this context  in order to take the existence of the bulk as a reality,  it is very illuminating to see clearly the bulk causality features in our model.  Specifically, we shall be interested in sending signals from one boundary to the other one for black holes/wormholes transition configuration, which would be a characteristic experiment to see the bulk causality.

This note is organized as follows. In section 2, we provide a brief summary of our setup and recall our bulk solution to the black holes/wormholes transition. In section 3, we describe how to send the signals from one boundary to the other one through the bulk and  what  the bulk causality tells us in this process. In section 4, we suggest how to probe the bulk by a certain experiment for two coupled boundary systems and what is the specific feature of this experiment, implying the bulk as a reality, compared to the usual direct interaction between them. In the final section, we conclude with some comments.




\section{ Bulk description of traversable wormholes
}\label{sec3}

In this section we summarize our bulk construction  of black hole/wormhole transition in~\cite{Bak:2018txn,Bak:2019mjd}. 
Our setup is based on two-dimensional Einstein-dilaton gravity, known as the JT model. It consists of a dilaton field $\phi$, a metric $g$, and  a free massive scalar field $\chi$ which does not couple directly the dilaton field. Explicitly, its action is given by
\begin{equation}
I=I_\textrm{top}+\frac1{16\pi G}\int_M d^2 x \sqrt{-g}\, 
\phi \left( R+\frac{2}{\ell^2}\right) + I_\textrm{surf} + I_M(g, \chi)\,,
\end{equation}
where 
\begin{align}
I_\textrm{top}&= \frac{\phi_0}{16\pi G}\int_M d^2 x \sqrt{-g}\, R\,,   \qquad 
I_\textrm{surf} = -\frac1{8\pi G}\int_{\partial M} 
	             \sqrt{-\gamma}\, (\phi_0 + \phi ) \, K \,, \notag \\
I_M &= -\frac12 \int_M d^2 x \sqrt{-g} ( \nabla \chi \cdot \nabla \chi + m^2
\chi^2 ) \,.
\end{align}
Here, $\ell$ is the AdS radius, and  $\gamma_{ij}$ and $K$ denote the induced metric and  
the extrinsic curvature at $\partial M$, respectively.
The dilaton field $\phi$ plays the role of the Lagrange multiplier and sets the metric to be
AdS$_2$ which can be written in the global coordinates as
\begin{equation}
ds^2 =\frac{\ell^2}{\cos^2 \mu} \left(-d\tau^2 + d\mu^2  \right)\,, \qquad \qquad  \mu \in \textstyle{[-\frac{\pi}{2},\frac{\pi}{2}]}\,.
\end{equation}
The metric variation leads to the equations of motion for the dilaton field $\phi$ as 
\begin{equation} \label{eqphi}
\nabla_a \nabla_b \phi -g_{ab} \nabla^2 \phi + g_{ab} \phi = -8 \pi G T_{ab}\,,
\end{equation}
where $T_{ab}$ is the stress tensor of the scalar field $\chi$,
\begin{equation}
T_{ab} = \nabla_a \chi \nabla_b \chi  -\frac{1}{2} g_{ab} 
        \left( \nabla \chi \cdot \nabla \chi + m^2 \chi^2 \right).
\end{equation}
In this note, we shall set $8\pi G =1$ for the simplicity.
With $T_{ab}=0$, the non-vanishing general vacuum solution for the dilaton field, which could be interpreted as black holes,
can be obtained in the form of 
\begin{equation} \label{dilaton}
\phi_{BH} = \bar\phi L
 \frac{\cos\tau }{ \cos \mu} \,,
\end{equation}
where we used $SL(2,{\bf R})$ to set the solution in the above form~\cite{Bak:2018txn}. Here, the length dimension parameter $L$ could be interpreted as the horizon radius, which can be shown as follows: By an appropriate coordinate transformation
the above AdS$_{2}$ metric can be set in the form of the black hole metric as
\begin{equation} \label{btz}
ds^2= - \frac{r^2-L^2}{\ell^2} dt^2+ \frac{\ell^2}{r^2-L^2} dr^2\,.
\end{equation}
We are in the low energy regime when  $L \ll \ell$. 

In the context of nearly AdS$_{2}$ gravity~\cite{Maldacena:2016upp,Almheiri:2014cka}, the degrees of freedom in this system reside on the boundaries and their dynamics may be described by a Schwarzian theory. In this description, boundary values of the metric and the dilaton could be taken in $\epsilon
\rightarrow 0$ limit as
\begin{equation} \label{bcondition}
	ds^2|_{\partial M} = -\frac1{\epsilon^2}d\tilde{u}^2\,, \qquad
	\phi|_{\partial M} = \frac{\ell \, \bar\phi }\epsilon\,,
\end{equation}
where $\tilde{u}$ denotes the (proper) boundary time. In order to construct the traversable wormholes in the bulk~\cite{Gao:2016bin}, we need the non-trivial scalar field $\chi$, whose boundary behavior is given by
\begin{align}   \label{}
\chi|_{\partial M } =  \epsilon^{\Delta} \tilde\alpha  + \cdots +   \epsilon^{1-\Delta} \tilde\beta  + \cdots\,, 
\end{align}
and the mixed boundary condition corresponds to 
\begin{equation} \label{}
\tilde\beta_{L/R} \propto \tilde\alpha_{R/L} \,,   
\end{equation}
where the subscript $L/R$ refers to the left/right boundary, respectively.
According to the AdS/CFT correspondence, this boundary  condition  corresponds to the double trace deformation of the boundary theory~\cite{Klebanov:1999tb,Witten:2001ua},  and may be written explicitly as the deformation of the boundary Hamiltonian 
\begin{equation} \label{defham}
\delta H(\tilde u) = -h(\tilde{u}){\cal O}_{R}(\tilde{u}){\cal O}_{L}(\tilde{ u})\,,
\end{equation}
where $\mathcal{O}_{R,L}$ are scalar operators of dimension $\Delta  \in (0, 1/2)$, dual to $\chi$. 
On the other hand, by taking $\tilde\beta$ as the source term, this deformation could be realized as  the non-local interaction with the $SL(2,{\bf R})$ gauge constraint (see \cite{Maldacena:2016upp,Almheiri:2014cka,Lin:2019qwu}). In this note, we take the L/R boundary proper time as the same one $\tilde{u}$. 
%
One may equivalently describe the system  by 
the Schwarzian derivatives at the boundaries with an interaction
term~\cite{Maldacena:2017axo,Maldacena:2018lmt},
\begin{equation} \label{sbdry}
S = \int d\tilde{u}\left[-\phi_{L}\, \textstyle{Sch} ( \tilde{u}) -\phi_{R}\,  \textstyle{Sch} (\tilde{u} )\right]  + S_{int}\,,
\end{equation}
where
\begin{equation} \label{}
\textstyle{Sch} ( \tilde{u}) \equiv  \Big\{ \tan \frac{\tau (\tilde{u})}{2},\tilde{u}\Big\}\,, \qquad \quad S_{int} \equiv  \frac{g}{2^{2\Delta}}\int d\tilde{u} \left[\frac{\tau'_{L}(\tilde{u})\tau'_{R}(\tilde{u})}{\cos^{2}\frac{\tau_{L}(\tilde{u})-\tau_{R}(\tilde{u})}{2}} \right]^\Delta\,.
\end{equation}
%
%
%
Here, $\phi_{L}=\phi_{R}$ can be identified with $\bar{\phi}$ in the bulk
and $g$ is a coupling proportional to the  parameter $h$, whose explicit identification is given by~\cite{Bak:2018txn}
\begin{equation}  \label{gtoh}
g = \frac{h}{2\pi}\frac{2^{2\Delta-1}\Gamma^{2}(\Delta)}{\Gamma(2\Delta)}.
\end{equation}
It has been shown~\cite{Maldacena:2016upp,Lin:2019qwu} that one may need to impose the appropriate gauge choice for the ghost-free dynamics in this Schwarzian description.


Though the metric is not affected by the presence of the matter field $\chi$,
the deformation \eqref{defham}  
via the scalar field $\chi$ affect 
the dilaton field $\phi$ through \eqref{eqphi}.
The general solution of \eqref{eqphi} can be written as \cite{Bak:2018txn}
\begin{equation}
	\phi = \phi_{hom}+ \varphi,
\end{equation}
where $\phi_{hom}$ denotes a vacuum solution ({\it i.e.} $T_{ab}=0$ case) and $\varphi$ is given by
\begin{equation} \label{DilDef}
\varphi(u,v) 
= \int^{u}_{u_{0}}dp \frac{\sin(p-u)\cos(p-v)}{\cos(u-v)} T_{uu}(p,v)\,.
\end{equation}
Here, $u$ and $v$ are global null coordinates defined by
$
u \equiv \frac{1}{2}(\tau + \mu)$, 
$v \equiv \frac{1}{2}(\tau - \mu)$.
Now, we consider eternal traversable wormholes (ETWs) in this setup.  
As was shown in 
\cite{Bak:2019mjd},  by turning on the double trace  interaction from the infinite past with the symmetric choice $\tau_{L} (\tilde{u}) = \tau_{R} (\tilde{u}) = \tau(\tilde{u})$ and by taking $\phi_{hom}=0$, 
 the ETWs could be realized by the dilaton field solution as
\begin{equation} \label{}
\phi_{\rm ETW}(u,v) = \frac{1}{2} \varphi_{\bf I} (u,v)\,,
\end{equation}
where 
\begin{equation} \label{varphiRI}
\varphi_{\bf I}(u,v) = {\textstyle \frac{4\ell^{2}\bar{h}
\Delta N_{\Delta}}{2^{2\Delta}}\frac{B(2-\Delta,2-\Delta)}{1-\Delta}
{\textstyle \frac{\sin^{3-2\Delta} |u-v| }{\cos |u-v| }} \, F\big(1-\Delta,1-\Delta\,;\, \frac{5}{2}-\Delta\,|\, {\textstyle \sin^{2} (u-v)} \big) }\,.
\end{equation}
Here, we would like to remind that the final expression of the dilaton field $\phi$ is obtained by a certain averaging procedure from a subregion contribution denoted as $\varphi_{\bf I}$. In this case, the sources at L/R boundaries are infinitely spread, and so one needs to add the subregion contribution appropriately. See~\cite{Bak:2019mjd} for the details and the notation. 
Whereas,  by turning on the double trace interaction at a certain time (concretely speaking, at $\tau_{i} =0$) and by adjusting its strength $h$ appropriately as  
\begin{equation} \label{}
 h = \bar\phi\frac{ 4\pi}{\Delta B(\Delta,\Delta)} \left(\frac{L}{\ell^2}\right)^{2(1-\Delta)}  \,,
\end{equation}
the ETW configuration can be joined with   the black hole configuration. Explicitly, the dilaton field can be obtained as 
\begin{equation} \label{bhtotw}
 \phi =\bar\phi \, L\,\,\frac{ \cos \tau }{ \cos \mu} +\varphi_{\rm TW}(u,v;q_{i}=\pi/4\,|\,L) \,,
\end{equation}
where we have taken $\phi_{hom}=\phi_{BH}$ and $\varphi_{\rm TW}$ denotes a specific combination of $\varphi_{\bf I}(u,v\,;\, q_{i})$ according to the value of $q_{i}$ (see Eq.(4.13) in~\cite{Bak:2019mjd}). Here, $\varphi_{\bf I}(u,v\,;\, q_{i})$ is given by
\begin{equation} \label{varphiRI}
\varphi_{\bf I} (u,v\,;\, q_{i})= \ell^{2}\frac{\bar{h}\Delta N_{\Delta}}{2^{2\Delta-2}} \int^{u}_{q_{i}} dq \Big[w^{1-\Delta}(1-w)^{2\Delta-1} + \Delta \frac{1+w}{1-w}B_{w}(1-\Delta,2\Delta)\Big]\,.
\end{equation}
Under our gauge choice, through the asymptotic expansion of the dilaton field $\phi$, it turns out that the boundary proper time coordinates $\tilde{u}$ is related to the bulk global time $\tau$ as 
\begin{equation} \label{bhtotw2}
 \tilde{u}=  \theta(-\tau)  \frac{\ell^2}{L} {\rm arctanh} \, \sin \tau+ \theta(\tau) \frac{\ell^2}{L}\tau \qquad    
  (\tau > -\pi/2)\,,
\end{equation}
which shows a transition from black holes to ETWs around the point $\tau=0$.  In the following, we consider sending signals on this black hole/ETW transition configuration. 
To see the bulk propagation of signal, we perturb the right system by turning on a source term to the total action
by
\be
L=L_{\rm TW}+ \gamma s(\tilde{u}) O^R_{\tilde{\Delta}}(\tilde{u}) \,,
\label{source}
\ee
where $L_{\rm TW}$ denotes the boundary action for two nearly CFT$_{1}$'s dual to the bulk gravity theory with the double trace deformation as
\be 
L_{\rm TW}= L_L + L_R + h\,  O^L_\Delta (\tilde{u})O^R_{\Delta}  (\tilde{u})\,.
\ee
One may recall that $L_{\rm TW}$ could be thought as the appropriate low energy (nearly CFT$_{1}$) limit of two coupled SYK models\footnote{The parameter $g$ in the Schwarzian theory could be written in terms of  the SYK model variables as $\bar\phi =\frac{\mu\alpha_{S}}{\cal J} $ and $\frac{g}{2^{2\Delta}}(\frac{N}{\phi_{R}})^{2\Delta-1}=\frac{\mu\alpha_{S}}{\cal J} \frac{c_{\Delta}}{(2\alpha_{S})^{2\Delta}}$ (see~\cite{Maldacena:2018lmt} for this identification).}~\cite{Maldacena:2017axo,Maldacena:2018lmt}. 

If the bulk starts with the two-sided black hole spacetime, the corresponding boundary system should 
be prepared at some initial time $\tu_{I}\ (< 0)$ in a particularly entangled state, more specifically, 
 a so-called thermofield double state.  Basically here we are following the general correspondence 
between the two-sided black hole spacetime and the thermofield double description of 
the L-R boundary system \cite{Maldacena:2001kr, Bak:2018txn}.  In this correspondence, 
the initial state is given by
\bea
|\Psi(\tu= \tu_I) \rangle  =\frac{1}{\sqrt{Z}} 
\sum_{n} \,  e^{-\left(\frac{\beta}{2}+ 2 i \tu_I \right) E_n }\, |n\rangle \otimes | n \rangle\,,
\label{initial}
\eea 
where $\sqrt{Z}$ is the normalization factor and the inverse temperature $\beta$ is an inverse of the Gibbons-Hawking temperature 
\be
\beta = \frac{1}{T}= \frac{2\pi \ell^2}{L}\,.
\ee 
Note that the L and R systems 
(without L-R interactions) possess an identical (nearly) CFT$_1$ Hamiltonian $H$ leading to identification 
$H_L  / \, H_R =  H \otimes 1  / \,1\otimes H $, respectively. $E_n$ and $|n\rangle $ are denoting eigenvalues and  eigenstates of the Hamiltonian $H$. Thus initially the left and right boundary systems are maximally entangled  for a given temperature $T$ and further the phase factor of each $ |n\rangle \otimes | n \rangle$ should be arranged initially as given in the above. 
The subsequent time evolution for the two-sided black hole system is given in terms of total Hamiltonian
$H_{\rm total}= H_L + H_R$. When an additional interaction is turned on, of course,  the total Hamiltonian of the system  becomes  $H_{\rm total}= H_L + H_R + H_{\rm int} $ with  $H_{\rm int}= \delta H$ in (\ref{defham}) {\it e.g.}  for the black hole/ETW transitional configuration.  When the transition occurs at $\tu_i =0$ as above, we note that the thermofield state at $\tu=0$ can be approximated to the ground state of the ETW system \cite{Maldacena:2018lmt}
and the subsequent evolution after the transition is almost stationary.

\section{Signaling in traversable wormhole }
In this section, we shall clarify signal propagations in our traversable wormhole systems. See~\cite{Hirano:2019ugo,Gao:2018yzk}  for some related aspects.  Of course, without the double trace deformation, the L-R systems are completely
 disconnected from each other and there is no way to send any signal 
from one side to the other. In the bulk of the black hole spacetime, a signal from one boundary enters the horizon and hits the singularity before meeting the other side. Namely the wormhole is not traversable in this case.

Once the L-R interaction is turned on, the wormhole becomes traversable. In this traversable wormhole 
system, there are generically two channels of the information propagation\footnote{We thank Juan Maldacena for the discussions and clarification on this point.}. One is the channel through the bulk. This respects the bulk causality and takes a certain amount of time for a signal transfer.  This can be used for the quantum teleportation 
sending quantum states through the bulk. The other channel is through the direct interaction which is 
instantaneous as the left and the right boundary times are identified and consequently interaction 
requires no time delay. 
Let us clarify these two channels respectively beginning with the bulk signaling. 

To this end, we turn on the source term on the right side (or the left side if one wishes) 
for an operator of dimension $\tilde\Delta$ as given  in (\ref{source}).
We shall switch on the source term only for $ \tilde{u} \in [\tilde{u}_i, \tu_f]$, so $s(\tu)$ vanishes when $\tu < \tu_i$ or $\tu > \tu_f$.
 First, let us recall that the bulk to boundary two point functions are given by
%
\begin{align}    \label{BBprop}
K_{L}(\tau- \tau_s,\mu; \Delta)&  =  {\cal N}_\Delta \Big[\frac{\cos\mu}{\cos(\tau-\tau_{s}) + \sin \mu} \Big]^{\Delta}\,, \nonumber \\
K_{R}(\tau- \tau_s,\mu,\Delta )&= {\cal N}_\Delta \Big[ \frac{\cos\mu}{\cos(\tau-\tau_{s}) - \sin \mu} \Big]^{\Delta} \,, 
\end{align}
where
\be 
{\cal N}_\Delta=  \frac{2^{\Delta-2}\Gamma^{2}(\Delta)}{\pi \Gamma(2\Delta)} \,.
\ee
%
%
Here, $L/R$  represent that the locations of the relevant boundary sources, which   are  taken, in global $(\tau,\mu)$ coordinates, as $(\tau_{s}, - \frac{\pi}{2})$ for  $K_{L}$  and as $(\tau_{s},  \frac{\pi}{2})$ for $K_{R}$, respectively. Since the exponent $\Delta$ is not an integer in our case, the phase of $K_{L/R}$ functions should be chosen appropriately, whenever the values inside brackets are negative. 

Since we turn on the right side source, we use $K_R$ to construct the corresponding bulk solution. The retarded condition requires
$\tau-\tau_s > -\mu +\frac{\pi}{2}$. The corresponding scalar field is solved by
\be 
\chi = \gamma 
\int^{\tau_f}_{\tau_i} d \tau_s\,  \tilde{s}(\tau_s) K(\tau- \tau_s,\mu; \tilde\Delta) \,,
\ee
where
\be 
\tilde{s}(\tau_s) =s(\tu(\tau_s)) (\tau'_s)^{\tilde{\Delta}-1}\,.
\ee
The retarded Green function $K$ is given by\footnote{We have taken the conventional  retarded Green function as 
$ {G_{\rm ret} (x,x') \equiv } i\langle  [\chi(x), \chi(x') ]\rangle\theta(\tau-\tau')$, { which satisfies  $(-\Box+m^{2})G_{\rm ret}(x) = \delta(x)$}. }
\bea  \label{RetG}
K(\tau,\mu; \Delta) &=& i\Big( 
K_R(\tau-i\epsilon,\mu; \Delta) -K^*_R(\tau+i\epsilon,\mu; \Delta) \Big)\theta\Big(\tau-\tau_s \Big)
\nonumber \\ 
&=&  2\sin\nu_R |K_R(\tau,\mu; \Delta) |  \theta\Big(\tau-\tau_s + \mu -\frac{\pi}{2} \Big)\,,
\eea
where $\nu_R$ refers to the phase  as in $K^{*}_{R}(\tau+i\epsilon) = e^{i\nu_{R}} |K^*_{R}(\tau + i\epsilon)|$.

{  At this stage, it would be useful to provide some details for the second equality in (\ref{RetG}).  This is related to the phase choice or the choice of branch cut and appropriate sheet in the multi-valued function  $f(x)=x^{\Delta}$ in the bulk to boundary function $K_{R}$.

If we ignore the multi-valuedness of the bulk to boundary function  $K_{R}$ in  (\ref{BBprop}),  it is $2\pi$ periodic with respect to $\tau-\tau_{s}$, and then the evaluation point $(u,v)$ would be space-like separated from  the source point $(\tau_{s},\frac{\pi}{2})$, as can be seen from Figure \ref{Phase1}. This naive periodicity give us the incorrect physical picture, since the evaluation point $(u,v)$ should be time-like separated from  the source point $(\tau_{s},\frac{\pi}{2})$.  
In order to rectify this  unwilling aspect, we need to introduce appropriate phase factors in $K_{R}$ (or $K^{*}_{R}$) in such a way that 
the evaluation point $(u,v)$ is time-like separated from the source point. In other words, we need to choose appropriate sheet for the correct value of the multi-valued function $f(x)=x^{\Delta}$. From the causality consideration, the appropriate phase choice in $K^{*}_{R}$ would be
\begin{equation} \label{}
\Big[\frac{\cos\mu}{\cos(\tau-\tau_{s}) - \sin \mu} \Big]^{\Delta} 
\rightarrow 
\Big[\negthinspace -\nth \frac{\cos\mu}{\cos(\tau-\tau_{s}) - \sin \mu} \Big]^{\Delta} = e^{i\pi\Delta}\Big[\frac{\cos\mu}{\cos(\tau-\tau_{s}) - \sin \mu} \Big]^{\Delta} \nth.
\end{equation}
The  additional phase  $e^{i\pi\Delta}$ should be taken into consideration whenever  we go across $+ 45^{\circ}$ red lines from yellow to green triangle regions in an upward direction and vice versa. 
Equipped with these additional phases one can obtain the final expression in (\ref{RetG}). 
\begin{figure}[thb]   
\begin{center}
\includegraphics[height=9.5cm]{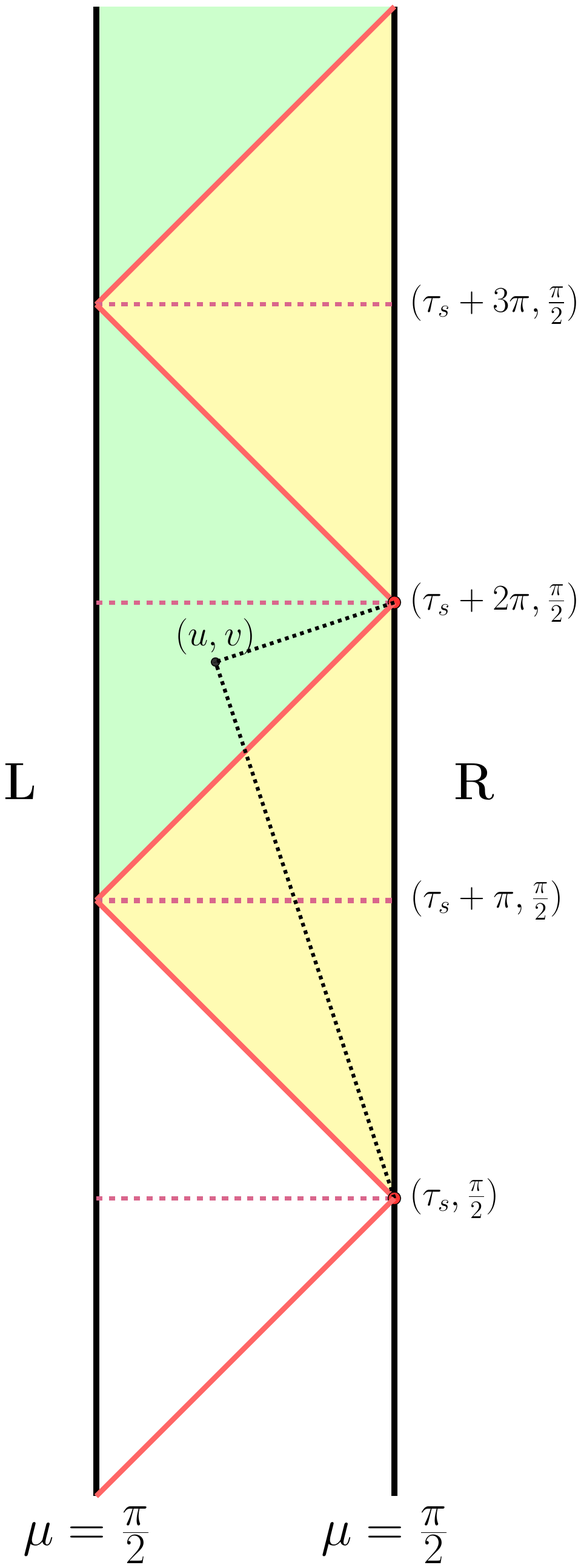} \includegraphics[height=9.5cm]{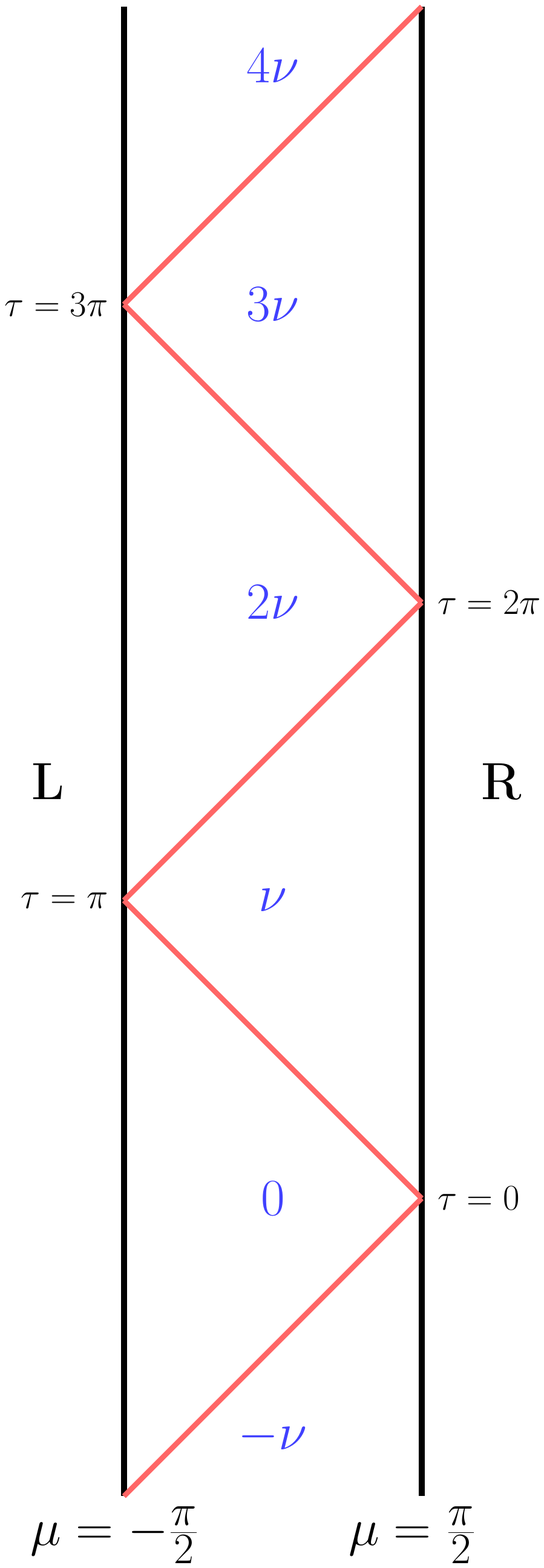}
\caption{\small On the left, the evaluation point $(u,v)$  should be timelike separated from  the source point $(\tau_{s}, \frac{\pi}{2})$, while the naive $2\pi$ periodicity implies that it is spacelike separated  from the source. 
 On the right, the phase assignment for $K^*_R$ is given where $\nu= \pi \Delta$  with $\tau_s=0$. }
\label{Phase1}
\end{center}
\end{figure} 
After these considerations, we obtain the phase assignments of  $\nu_R$ in Figure~\ref{Phase1} as follows\footnote{This phase assignment is already made  in our previous work~\cite{Bak:2019mjd}.}:
}
\bea
\nu_R(\tau-\tau_s,\mu;\Delta) =\left\{ 
\begin{array}{ccc}
(2n+1)\pi {\Delta} & {\rm if } &     -\mu+\big(2n+\frac{1}{2}\big)\pi  <  \tau-\tau_s < \mu +\big(2n+\frac{3}{2}\big)\pi  \\
2n\pi {\Delta} & {\rm if }  & \ \  \  \,  \mu+ \big(2n-\frac{1}{2}\big)\pi  <  \tau-\tau_s <   - \mu +\big(2n+\frac{1}{2}\big)\pi  
\end{array} \,.
\right. \nonumber
\eea
We illustrate this phase assignment  on the right of  Figure \ref{Phase1}.


In the asymptotic region of  the left boundary $\mu\sim   -\pi/2$, the leading order of the scalar field behaves as 
\be
\chi = O(\cos^{\tilde\Delta} \mu) \,.
\ee 
Hence, no source term is present in the left side while the expectation value of operator $O^{\tD}_L$ is induced by the bulk transmission. The induced expectation value is given by
\be 
\langle O_L(\tau)\rangle =  2 {\cal N}_{\tilde\Delta} \gamma
\int^{\tau_f}_{\tau_i} d \tau_s\,  \tilde{s}(\tau_s) \, \Big[\frac{1}{1+ \cos(\tau-\tau_{s}) } \Big]^{\tD}
\theta\big(\tau-\tau_s - \pi \big) \sin \nu_{LR}
\ee
with ${\nu}_{LR}$ denoting $\nu_R(\tau-\tau_s;-\pi/2;\tilde{\Delta})$. The factor in the square bracket becomes infinity when
$\tau-\tau_s=(2n+1)\pi $ with integer $n$. Therefore, the information on the left side appears only when $\tau > \tau_{s} +\pi$ and the corresponding propagation respects bulk causality.
 Here, we also record the expectation value induced on the right side given by
\be 
\langle O_R(\tau)\rangle =  2 {\cal N}_{\tilde\Delta} \gamma
\int^{\tau_f}_{\tau_i} d \tau_s\,  \tilde{s}(\tau_s) \, \Big[\frac{1}{1 - \cos(\tau-\tau_{s}) } \Big]^{\tD}
\theta\big(\tau-\tau_s  \big) \sin \nu_{RR}
\ee
with ${\nu}_{RR}$ denoting $\nu_R(\tau-\tau_s; \pi/2;\tilde{\Delta})$. 
Note that the factor in the square bracket in this RR case becomes infinity 
when $\tau-\tau_s=2n\pi  $ with integer $n$.
 

\begin{figure}[th!]

\centering  
\includegraphics[height=6.5cm]{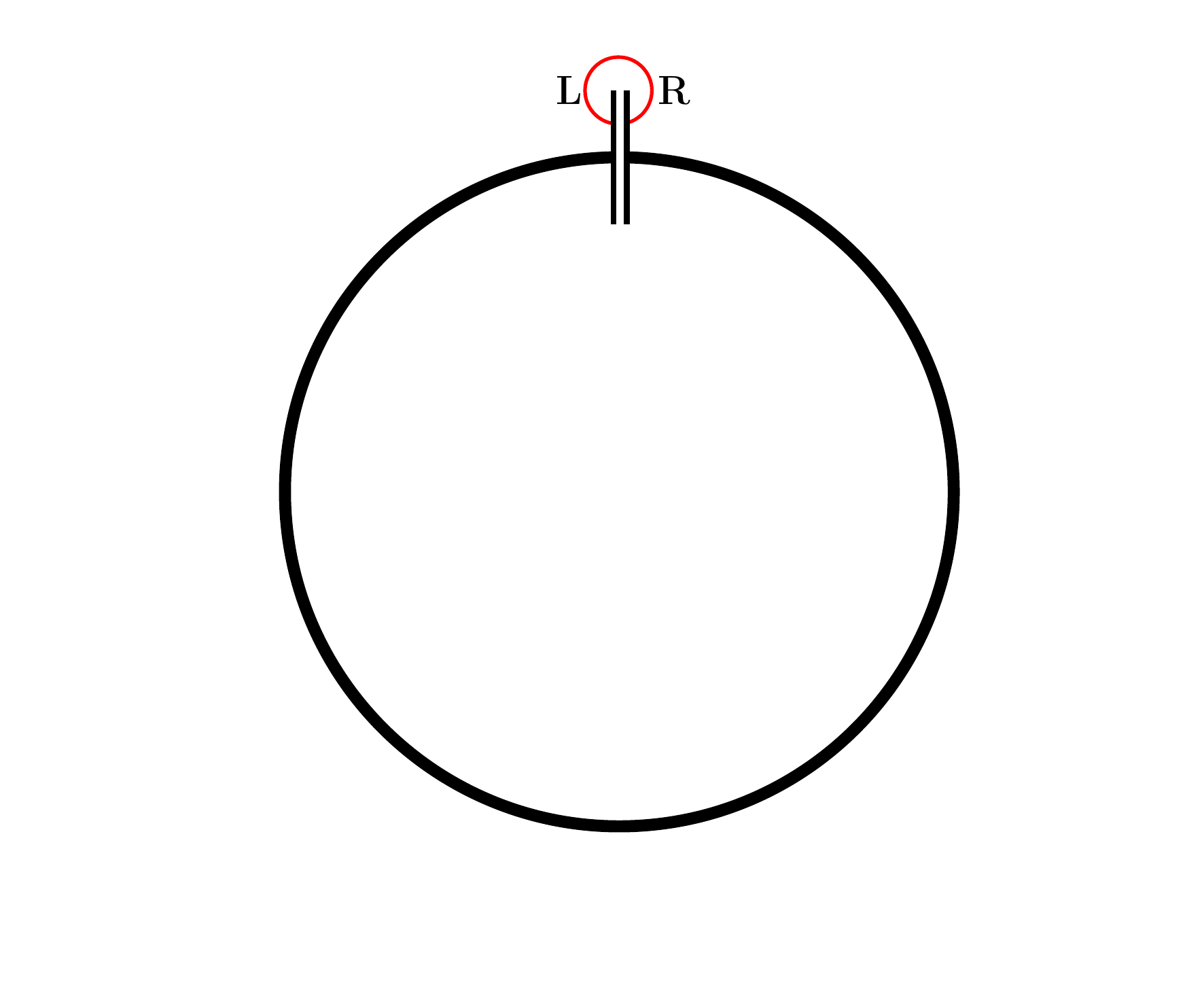}
\vskip-.5cm
\caption{\small       The spatial part of the Lorentzian traversable  wormhole system is depicted schematically. One channel of signal propagation
is through the bulk, which takes a certain amount of time due to the bulk causality. The other is the direct propagation through the interaction between the L-R systems, which is instantaneous.} 
\label{figschwh}
\end{figure}

 One may perform the corresponding CFT analysis. There the propagator from the right source to the expectation value on the left is simply given by
 \be
k_{LR}(\tau- \tau_s; \tilde\Delta)   =    2 {\cal N}_{\tilde\Delta} \Big[\frac{1}{1+ \cos(\tau-\tau_{s}) }\, \Big]^{\tilde\Delta}
 \theta\big(\tau-\tau_s-\pi \big)\sin \nu_{LR} \,, 
 \ee 
which is following from the nearly conformal symmetry of the traversable wormhole system. 
 Then from this the above result can be regained straightforwardly. 


We now turn to the discussion of  the other channel, which may be more clearly seen 
 in the Schwarzian description in (\ref{sbdry}). 
 In this system, we consider a perturbation on the right side as $\delta \tau_{R}$\footnote{In the SYK model, $\tau_{R}$ and $\tau_{L}$ corresponds to the reparametrization modes of L/R system which are the soft modes of the nearly CFT systems.}.
The left side $\tau_L$  will then be instantaneously affected by the perturbation of $\tau_R$, which is basically due to
the L-R interaction term,       as may be seen as follows.  By taking into account the linear perturbation of the Schwarzian action in Eq.~(\ref{sbdry}), one can obtain  
\begin{equation} \label{}
\frac{2}{\tau'^{4}_{g}}\, \delta \tau''''_{L} +  \frac{2 (2-\Delta )}{\tau'^{2}_{g}}\, \delta \tau''_{L}  + \delta \tau_{L} = \frac{2\Delta}{\tau'^{2}_{g}}\,  \delta \tau''_{R} + \delta \tau_{R}\,, \qquad {}'\equiv \frac{d}{d\tilde{u}}\,,
\end{equation}
where $\tau'_{g}$ denotes the constant  corresponding to the eternal traversable wormhole solution chosen\footnote{The constant $\tau'_{g}$ value could be related to $\beta_{g}$, which would be introduced later  in Eq.~(\ref{betag}), as $\tau'_{g} = \frac{2\pi}{\beta_{g}}$. Note also that there is an additional equation, in which $\tau_{L}$ and $\tau_{R}$ are interchanged on the linear perturbation of the Schwarzian action.} as  $\tau'_{L}=\tau'_{R}=\tau'_{g}$. 
This shows us the instantaneous reaction of $\tau_{L}$ by  the change of  $\tau_{R}$ in the amount of $\delta \tau_{R}$.


   These two channels are schematically depicted in Figure  \ref{figschwh}. The big circle represents the bulk traversable   wormhole and small red circle represents the instantaneous boundary channel directly through the interactions. Our wormhole system can be embedded into higher dimensions \cite{Maldacena:2018gjk} (see also~\cite{Freivogel:2019lej}).     In this case, the L and R systems can be  spatially separated in the boundary spacetime.    Of course, the boundary signal propagation through the boundary interactions takes some 
amount of time depending on the separation. Then the overall causality requires that the bulk propagation
through the wormhole should not be faster than the boundary propagation, as dictated by the causality of the boundary theory.   Interestingly, this is consistent with the Gao-Wald theorem~\cite{Gao:2000ga,Kelly:2014mra}, even though some conditions for the theorem are violated in our setup.
   
    In the next section, we shall discuss a possible experimental probe of the bulk channel purely in terms of  boundary    operations.   This discussion is focused on the characteristics of the bulk channel that can also be  used for the quantum teleportation.

\section{Experimental probes of wormholes}

In the L-R boundary systems, suppose that signals are sent from the R system 
and the observer in the L system detects them. The observer may 
wonder from which channel they emerged. The boundary channel would always
be an obvious candidate since, without any boundary interaction, the 
two systems are completely disconnected. Then it would be more interesting to
ask whether there is any situation that the L observer would naturally
conclude that the detected signal must have passed not through the usual
boundary channel but through other channel, {\it i.e.}, the bulk channel. 
Note that from the pure boundary viewpoint, the existence of the bulk itself
is nontrivial and also a boundary interaction does not always guarantee
traversability of the bulk channel.

\begin{figure}[th!]
\centering  
\includegraphics[height=8cm]{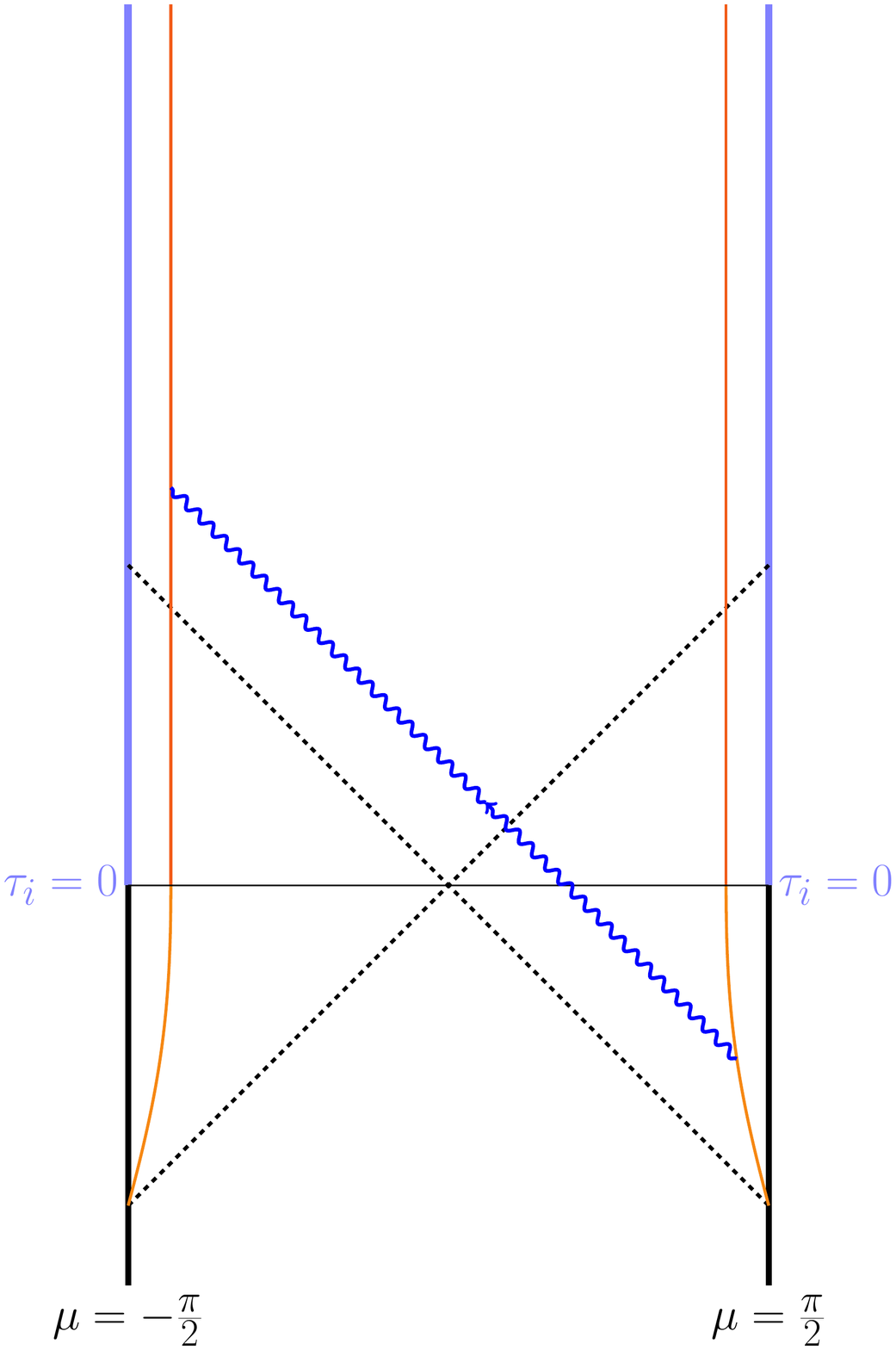}\includegraphics[height=7.35cm]{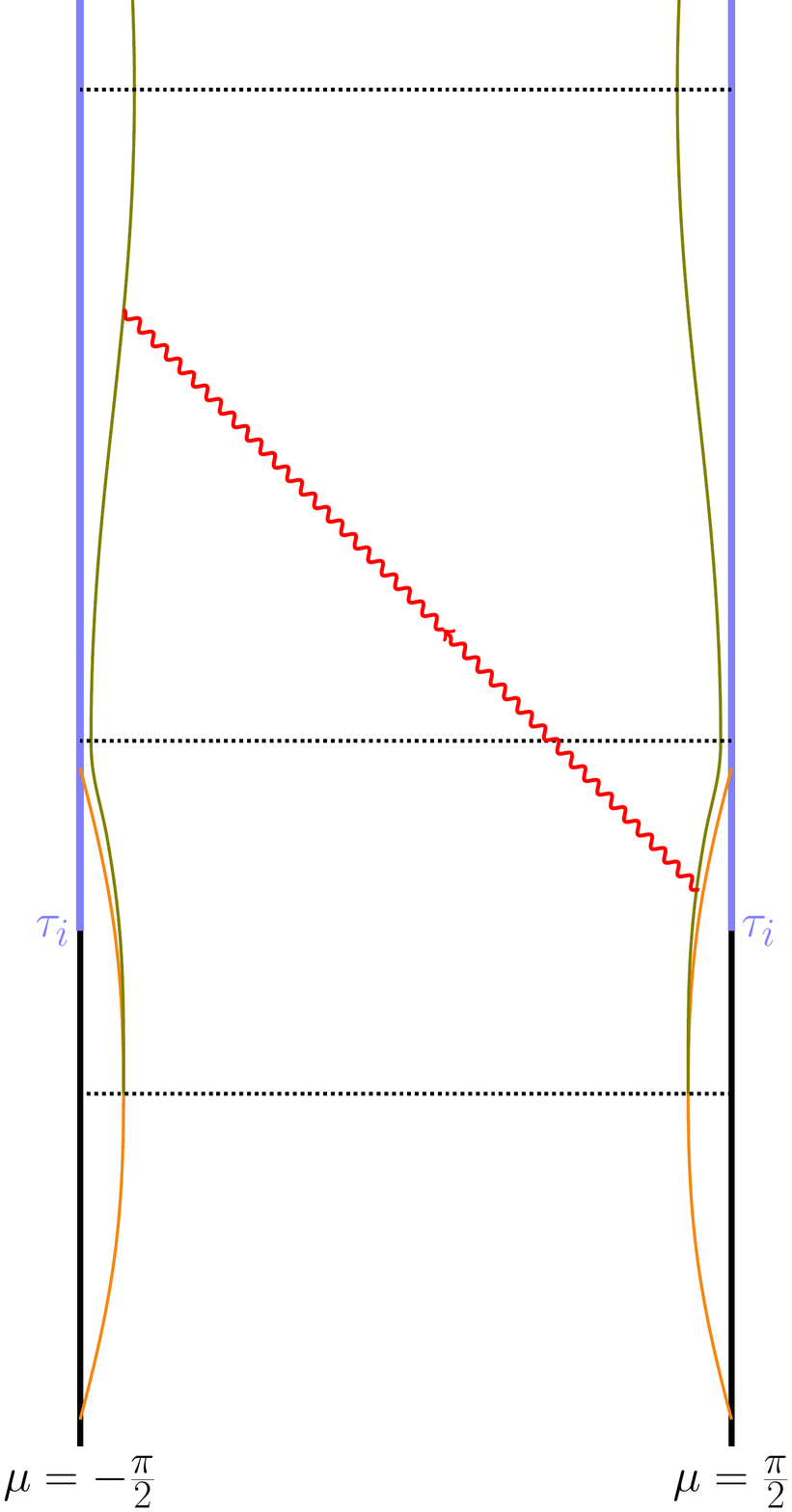}
\caption{\small On the left we depict a transitional configuration of the black hole to ETW where
	the L-R boundary interaction is turned on at $\tilde u_i = 0$. On the right we depict a transitional configuration of the black hole to an excited (oscillatory)  ETW state where
	the L-R boundary interaction is turned on at $ \tu_i > 0$. 
	Signals sent before turning on the L-R interaction through
	the bulk channel emerges to the other side after some amount
	of time with modulated frequencies,
	which is a clear evidence of the existence of the bulk channel
	and traversability.  
} 
\label{figsignal}
\end{figure}

To answer this question, let us consider the configuration \eqref{bhtotw},
which describes the black hole/ETW transition as depicted on the left panel of  Figure~\ref{figsignal}, dubbed
as transparentization of a black hole to an ETW~\cite{Bak:2019mjd}.
The boundary time $\tilde u$ is then related to the bulk global
time as \eqref{bhtotw2}. In this configuration, one may send a signal
through the bulk from one to the other side with perturbation \eqref{source}.
The boundary time $\tilde u_o$ that the signal sent at $\tilde u_s$ is
observed at the other side is given by
\begin{equation}  \label{SignD}
\tilde {u}_o = \frac\beta2 + \theta(\tilde u_s) \tilde u_{s}
+ \frac\beta{2\pi} \theta(-\tilde u_s) \arcsin \tanh \frac{2\pi\tilde u_s}\beta.
\end{equation}
As discussed in~\cite{Bak:2019mjd}, the signals sent during 
$-\infty < \tilde u_s < 0 $, 
which is before turning on the boundary interaction,
would come out at the other side during the time 
$\frac\beta4 < \tilde u_o < \frac\beta2$, obeying the bulk causality. 
Moreover, they are blue-shifted.
If the signals are sent with frequency $\omega_s$ which 
is controlled by the source term $s(\tilde u)$ in \eqref{source},
the frequency $\omega_o$ observed at the other side is given by
\begin{equation}
	\omega_o = \omega_s \cosh \frac{2\pi \tilde u_s}\beta.
\end{equation}
Note that the signal sent earlier would be blue-shifted more strongly.
For example, the blue-shift factor is 268 for $\tilde u_s=-\beta$, while
it is as high as 143376 for $\tilde u_s = -2\beta$.

On the other hand, it is obvious that signals from the boundary channel 
would emerge to the other side instantaneously without any blue-shift factor. 
Therefore, even though there are two channels 
of signal propagation, they could be clearly distinguished in the black 
hole/ETW transitional configuration. 

%
\begin{figure}[htbp]   
\vskip-3cm
\begin{center}
\includegraphics[width=0.45\textwidth]{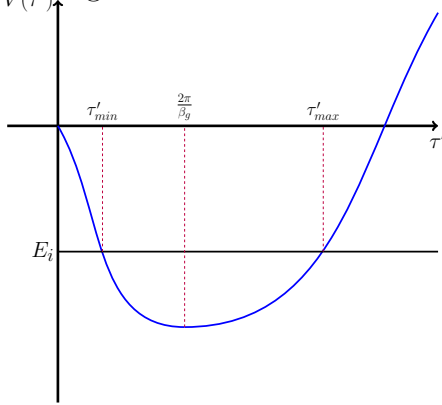}
\vskip-1cm
\caption{The potential as a function of $\tau'$ is depicted in this figure. We also indicate the bounded motion and its energy which is negative. }
\label{potgraph}
\end{center}
\end{figure} 

It might be the case that experimentally realizing the particular set of phases appearing in the 
thermofield initial state  in (\ref{initial}) is not that straightforward. To avoid such complications, one may turn 
on the L-R interaction at $\tu =\tu_i >0$ and consider the system from $\tu_I =0$ in which the initial state
does not involve any such phases. The corresponding Schwarzian dynamics is described by
\begin{equation} \label{SchwR}
L= \bar{\phi} \, \Big( \frac{\tau''}{\tau'}\Big)^2 -\bar{\phi} \, (\tau')^2 +
\frac{g}{2^{2\Delta}} (\tau')^{2\Delta} \theta(\tu -\tu_i)\,,
\end{equation}
where we set $\tau_L = \tau_R= \tau$ as before\footnote{Normally we take ${\cal Y}=\log \tau'$ as our dynamical
variable and then the kinetic term is given by $K= \bar\phi \, ({\cal Y}')^2 $ together with the potential  
$V=
 \bar{\phi} \,e^{2{\cal Y}} -
\frac{g}{2^{2\Delta}} e^{2\Delta {\cal Y}} \theta(\tu -\tu_i)$.}. We begin with a black hole with an inverse
temperature $\beta$ described by a solution 
$\sin \tau = \tanh \frac{2\pi \tu}{\beta}$
for $\tu < \tu_i$. At $\tu=\tu_i$ where the interaction is turned on, $\tau'$ and  $\tau''$ should be 
continuous. We require $E_i <0$ right after the interaction is turned on in order to have a bound state motion ($\tau' > 0$) leading to a traversable wormhole state. This then gives a condition 
\be 
\Delta \cosh^{2\Delta}  \frac{2\pi \tu_i}{\beta} < \left(\frac{\beta}{\beta_g}\right)^{2(1-\Delta)}\,,
\ee 
where we introduce a length scale $\beta_g$ by
\be \label{betag}
\beta_g =2\pi \left(\frac{g\Delta}{2^{2\Delta}\bar\phi}\right)^{-\frac{1}{2(1-\Delta)}} \,.
\ee
The subsequent time evolution with variable $\tau'$  satisfies the  
equation of motion 
\be 
E_i =  \bar{\phi}\left[ \Big( \frac{\tau''}{\tau'}\Big)^2 +\bar{\phi} \, (\tau')^2 \right] -
\frac{g}{2^{2\Delta}} (\tau')^{2\Delta} \,,
\ee
as depicted in Figure~\ref{potgraph}. This wormhole configuration is  in an excited state
of the ETW system and becomes oscillatory as depicted on the right side panel of Figure \ref{figsignal}.  
 Note that the L/R boundary trajectories (green curves in the right panel in Figure~\ref{figsignal})  are  determined by the equations 
\begin{equation} \label{}
\mu (\tilde{u})  = \mp\Big[ \frac{\pi}{2} - \epsilon \tau'(\tilde{u}) + {\cal O}(\epsilon^{2}) \Big]\,.
\end{equation}
%

We now send a signal from the right at $\tu = \tu_s $ with $0 < \tu_s $.  
One 
finds the signal through the wormhole appears on the left side when $\tu_o$ determined by the condition
$\tau(\tu_o) =\tau(\tu_s) +\pi$. Thus again there is the corresponding delay and the frequency modulation of signal  is given by
\be 
\omega_0 = \omega_s \frac{\tau'(\tu_o)}{\tau'(\tu_s)}\,,
\ee
which may be either red-shifted or blue-shifted depending on $\tu_s$ and the parameter $\beta_g$.  In Appendices \ref{AppA}  and \ref{AppB}, we 
present  some detailed analysis of this bulk signaling problem
 for two specific cases  where  the L-R interaction is turned on at $\tilde{u}_{i} >0$.

Finally, one may also consider the ETW/black hole transitional configuration 
given in \cite{Bak:2019mjd} where its transition is now induced by turning off the L-R interaction at 
$\tu_i=0$. In this case, one finds again a delay together with a modulation of the signal frequency.
Its experimental realization is expected to be relatively simple since our system in this case starts in the ground state 
of the ETW system which is stationary until the transition.

%

In condensed matter physics, there are several attempts to engineer 
the SYK model with ultracold gases, graphene flakes, quantum wires 
or 3D topological insulators (see {\it e.g.} \cite{Franz:2018cqi} and 
references therein). Then it might be possible to experimentally realize 
the holographic dual of the traversable wormhole system by entangling 
two SYK model systems. By turning on or off different type of interactions 
between them, one may send signals from one side to the other. If
frequency-modulated signals are detected after some amount of time delay for a
double trace interaction while no such signals are observed for
other cases, it could be considered as a strong evidence that 
the bulk is real and a traversable wormhole is formed between the two
systems.

\section{Conclusions}
 In this note, we clarify the nature of signal propagation in traversable wormhole system.
 Basically we show that there are two independent channels. One is through the bulk wormhole
 and the other is the direct boundary channel which is via the boundary interaction. We also 
describe in detail 
how the information can be transferred from one side to the other through the bulk wormhole using 
a bulk scalar field dual to a boundary scalar operator.

We then take the examples of the bulk  representing the transition from the black hole 
from/to the ETW geometry where the transition is induced by turning off/on the double trace deformation.   
The signal from one side emerges to the other side after certain 
amount of time perfectly respecting the bulk causality. 
The signal frequency detected on the other side will be 
in general highly modulated 
 while passing through the bulk of traversable wormhole. 
 These two facts may be 
used to show the existence of the bulk wormhole since the frequency-modulation  patterns as well as 
the time delay imprint clear characteristics of the bulk geometry. Also note that, in this measurement, 
one needs to perform a simple set of  purely boundary operations only. 

The experimental realization of our setup in the boundary system seems feasible without any 
fundamental difficulties such as probing the behind horizon degrees in ordinary two-sided black hole systems.

\subsection*{Acknowledgement}
We thank Juan Maldacena for enlightening discussions for clarifying the nature of interactions in 
traversable wormhole systems.
DB was
supported in part by
NRF Grant 2017R1A2B4003095 and by Basic Science Research Program
through National Research Foundation funded by the Ministry of Education
(2018R1A6A1A06024977). C.K.\ was supported by NRF Grant 2019R1F1A1059220.
S.-H.Y. was supported by 
NRF Grant  
2018R1D1A1A09082212.

\appendix 

\section{Perturbative solution to ETW   } \label{AppA}
\renewcommand{\theequation}{A.\arabic{equation}}
  \setcounter{equation}{0}
In this appendix we provide some details on the signal sending on the configuration in the right panel of Figure~\ref{figsignal}. 
This configuration could be described by adjoining  the black hole solution for $\tilde{u} \le \tilde{u}_{i}$ to a traversable wormhole solution for $\tilde{u} > \tilde{u}_{i}$. In the context of the Schwarzian dynamics in 
(\ref{SchwR}), we consider a perturbative solution around the ETW solution $\tau' = \frac{2\pi}{\beta_{g}}$ (see Figure~\ref{potgraph})  for $\tilde{u} > \tilde{u}_{i}$,  by taking 
\begin{equation} \label{}
\tau' (\tilde{u}) =  \frac{2\pi}{\beta_{g}} + \delta \tau' (\tilde{u}) \nonumber 
\end{equation}
with the Lagrangian in (\ref{SchwR}) as
\begin{equation} \label{}
L=   \bar{\phi}\Big(\frac{2\pi}{\beta_{g}}\Big)^{2} \frac{1-\Delta}{\Delta}+
 \bar{\phi}\Big(\frac{\beta_{g}}{2\pi}\Big)^{2}\Big[ (\delta\tau'')^{2} - \omega^{2}_{g}(\delta \tau')^{2} +\cdots \Big] \quad  \text{for} ~~ \tilde{u} > \tilde{u}_{i}\,.
\end{equation}
Here, $\omega_{g}$ is defined by
\begin{equation} \label{}
 \omega_{g} \equiv \frac{2\pi}{\beta_{g}} \sqrt{2(1-\Delta)}\,.
\end{equation}

To match the perturbative solution for $\tilde{u} > \tilde{u}_{i}$ with the black hole solution for $\tilde{u} \le \tilde{u}_{i}$ at $\tilde{u} = \tilde{u}_{i}$, we take the parameters in such a way that  $\frac{2\pi}{\beta}|\tilde{u}_{i}| \ll 1$ and $\big|\frac{\beta}{\beta}_{g} -1\big| \ll 1$. Then, the matched solution is given by
\begin{equation} \label{}
\tau'(\tilde{u}) = %
\left\{ \begin{array}{lll}    \frac{2\pi}{\beta}\, \frac{1}{\cosh\frac{2\pi}{\beta} \tilde{u}} 
&   \quad \text{for} &     \tilde{u} \le \tilde{u}_{i} \\
\frac{2\pi}{\beta_{g}} - \frac{2\pi}{\beta_{g}}  \big[a\cos\omega_{g}\tilde{u} + (\frac{2\pi }{\beta_{g}})^{2} \frac{\tilde{u}_{i}}{\omega_{g}} \sin\omega_{g}\tilde{u} \big]  + \cdots & \quad \text{for}  &    \tilde{u} >\tilde{u}_{i}
\end{array}  \right.   \,,
\end{equation}
where  $a$ is  defined by $a\equiv \frac{\beta}{\beta_{g}} -1$ and   $\cdots$ denotes higher order terms in $a$ and/or  $\frac{2\pi}{\beta}\tilde{u}_{i}$.

Now, let us consider a signal which is sent at $\tilde{u} = \tilde{u}_{s}$ in such a way that $|\tilde{u}_{s}|  < |\tilde{u}_{i}|$.  Since $ \frac{2\pi}{\beta} |\tilde{u}_{s}| <  \frac{2\pi}{\beta}  |\tilde{u}_{i} | \ll 1 $, one can see that 
\begin{equation} \label{}
\tau(\tilde{u}_{s}) = \frac{2\pi}{\beta} \tilde{u}_{s}\,.
\end{equation}
Let this signal be observed at $\tilde{u}_{o}$, which is realized in the bulk global time  as $\tau(\tilde{u}_{o}) =  \pi + \tau(\tilde{u}_{s})$. By using the perturbative solution for $\tilde{u} > \tilde{ u}_{i}$, one can also see that 
\begin{equation} \label{}
\tau(\tilde{u}_{o}) = \frac{2\pi}{\beta_{g}} \tilde{u}_{o} - \frac{2\pi}{\beta_{g} \omega_{g} }  \Big[a\sin\omega_{g}\tilde{u}_o - \Big(\frac{2\pi }{\beta_{g}}\Big)^{2} \frac{\tilde{u}_{i}}{\omega_{g}} ( \cos\omega_{g}\tilde{u}_o -1) \Big]  + \cdots\,, \nonumber
\end{equation}
where we have chosen the initial condition as $\tau(\tilde{u}=0)=0$.   
As a result, one obtains 
\begin{align}    \label{}
\tilde{u}_{o} - \tilde{u}_{s} &= \frac{\beta_{g}}{2}  + \frac{a}{\omega_{g}} \sin \frac{\beta_{g}\omega_{g}}{2} +2  \Big(\frac{2\pi }{\beta_{g}}\Big)^{2} \frac{\tilde{u}_{i}}{\omega^{2}_{g}}  \sin^{2} \frac{\beta_{g}\omega_{g}}{4}   + \cdots \,,  \\
\omega_{o}  
 & =  \omega_{s}  \Big[ 1 + 2a \sin^{2}\frac{\beta_{g}\omega_{g}}{4}  - \Big(\frac{2\pi }{\beta_{g}}\Big)^{2} \frac{\tilde{u}_{i}}{\omega_{g}} \sin\frac{\beta_{g}\omega_{g}}{2}   +\cdots   \Big]\,,  
\end{align}
where $\omega_{s}$ and $\omega_{o}$ are the frequencies at $\tilde{u}_{s}$ and $\tilde{u}_{o}$, respectively.  
For purposes of comparison, one may notice   that the result  in (\ref{SignD})   becomes, under the present assumption that $\frac{2\pi}{\beta}\tilde{u}_{s} \ll 1$, 
\begin{equation} \label{}
\tilde{u}_{o} = \frac{\beta}{2} + \tilde{u}_{s}\,,
\end{equation}
which is consistent with the above result up to the relevant order. 
This result shows us clearly that the frequency modulation of the signal  could be either blue-shifted or red-shifted according to the parameter choice.   
%

\section{$\Delta=\frac{1}{2}$ case } \label{AppB}
\renewcommand{\theequation}{B.\arabic{equation}}
  \setcounter{equation}{0}
In this appendix we consider a special case of $\Delta =\frac{1}{2}$ in (\ref{SchwR}). In this case the bounded motion for $\tilde{u} > \tilde{u}_{i}$ would have a conserved energy as 
\begin{equation} \label{}
H =  \bar{\phi} \Big[ \Big( \frac{\tau''}{\tau'}\Big)^2 + \tau'^2 - 
\frac{g}{2\bar{\phi}}\, \tau'  \Big] = E_{i} \,,
\end{equation}
which gives us 
\begin{equation} \label{}
\frac{\tau''}{\tau'} = \pm \sqrt{(\tau' - \tau'_{min})(\tau'_{max}-\tau')}\,.
\end{equation}
Here, $\tau'_{min/max}$ denotes the turning points of the bounded motion (see Figure~\ref{potgraph}) and are related to the parameters as 
\begin{equation} \label{}
\frac{1}{2}(\tau'_{min} + \tau'_{max} ) = \frac{2\pi}{\beta_{g}}=\frac{g}{4\bar{\phi}}\,, \qquad  \bar{\phi} \, \tau'_{min}\,  \tau'_{max} = -E_{i} \,.
\end{equation}
Integrating the above differential equation, 
one can see  that the solution for the bounded motion is given by
\begin{equation} \label{}
 \tau'(\tilde{u})  = \frac{2\tau'_{max}\tau'_{min}}{[\tau'_{max}+\tau'_{min}- (\tau'_{max}- \tau'_{min})\sin  \sqrt{\tau'_{max}\tau'_{min}} (\tilde{u}- \tilde{u}_{c}) ]}\,,
\end{equation}
where $\tilde{u}_{c}$ is an integration constant.  By integrating this expression, 
%
one obtains  finally 
\begin{equation} \label{}
\tan \Big[\frac{\tau -\tau_{c}}{2} \Big] =\frac{\tau'_{max} + \tau'_{min}}{2\sqrt{\tau'_{max}\tau'_{min}}} \tan\Big[ \frac{\sqrt{\tau'_{max}\tau'_{min}}}{2} (\tilde{u}-\tilde{u}_{c}) \Big]  - \frac{\tau'_{max}- \tau'_{min}}{2\sqrt{\tau'_{max}\tau'_{min}}}\,,
\end{equation}
where $\tau_{c}$ is another integration constant.

As before, one can fix the constants by matching the solution to the black hole solution at $\tilde{u}=\tilde{u}_{i}$ by using the continuity of $\tau'$ and $\tau''$.  And then, it is straightforward to consider the time delay and frequency modulation for the signal sent at $\tilde{u}= \tilde{u}_{s}$, at least numerically.


\end{document}